\documentclass[aps,prl,twocolumn,secnumarabic,balancelastpage,amsmath,amssymb,floatfix,showpacs,longbibliography,reprint,superscriptaddress]{revtex4-1}
\usepackage{graphicx} 
\usepackage{color}      
\usepackage{hyperref} 

\begin{document}

\title{Driven-dissipative many-body systems with mixed power-law interactions: Bistabilities and temperature-driven non-equilibrium phase transitions}

\author{Nikola \v{S}ibali\'c}
\email{nikolasibalic@physics.org}
\affiliation{Joint Quantum Center (JQC) Durham-Newcastle, Department of Physics, Durham University, South Road, Durham, DH1 3LE, United Kingdom}
\author{Christopher G. Wade}
\affiliation{Joint Quantum Center (JQC) Durham-Newcastle, Department of Physics, Durham University, South Road, Durham, DH1 3LE, United Kingdom}
\author{Charles S. Adams}
\affiliation{Joint Quantum Center (JQC) Durham-Newcastle, Department of Physics, Durham University, South Road, Durham, DH1 3LE, United Kingdom}
\author{Kevin J. Weatherill}
\affiliation{Joint Quantum Center (JQC) Durham-Newcastle, Department of Physics, Durham University, South Road, Durham, DH1 3LE, United Kingdom}
\author{Thomas Pohl}
\affiliation{Max Planck Institute for the Physics of Complex Systems, D-01187 Dresden, Germany}

\date{\today}

\begin{abstract}

We investigate the non-equilibrium dynamics of a driven-dissipative spin ensemble with competing power-law interactions. We demonstrate that dynamical phase transitions as well as bistabilities can emerge for asymptotic van der Waals interactions, but critically rely on the presence of a slower decaying potential-core. Upon introducing random particle motion, we show that a finite gas temperature can drive a phase transition with regards to the spin degree of freedom and eventually leads to mean-field behaviour in the high-temperature limit. Our work reconciles contrasting observations of recent experiments with Rydberg atoms in the cold-gas and hot-vapour domain, and introduces an efficient theoretical framework in the latter regime.
\end{abstract}


\maketitle



The idea that matter rapidly relaxes towards a thermal ensemble \cite{gibbs} has led to a profound understanding of many macroscopic phenomena within the powerful framework of equilibrium statistical physics. More recently, the experimental success in realizing synthetic many-body systems with controllable dissipation has motivated broad explorations of non-thermal steady states \cite{sieberer15}. Examples include cold atoms in cavities \cite{ritsch13}, semiconductor exciton-polariton condensates \cite{carusotto13}, trapped ion crystals \cite{britton12} and laser-driven Rydberg gases \cite{Saffman2010}. The interplay of coherent and dissipative dynamics in such driven-dissipative systems generates non-equilibrium phases and transitions that may have no equilibrium equivalent. An evident example of such distinct behaviour is the emergence of multiple steady states.

Signatures of bistable many-body phases and hysteretic behaviour are reported in experiments on cold \cite{Malossi2014} and thermal \cite{Carr2013b} Rydberg gases, which offer controllable particle interactions, dissipation and coherent driving. While the basic physics suggests a conceptually simple description in terms of a dissipative spin ensemble \cite{Lee2011a,lukin13,Lee2012,Hu2013,honing13,Marcuzzi2014c,Ates2012,Lesanovsky2014a,Weimer2015a,Weimer2015b,Maghrebia,Hoening2014a,Mendoza-Arenas2015}, understanding its many-body dynamics has proved challenging. Lattice mean field (MF) descriptions \cite{Hu2013,Marcuzzi2014c}, for instance, relate cold atom experiments \cite{Malossi2014} to the formation of a bistable steady state, while variational calculations \cite{Weimer2015a,Weimer2015b} suggest an interpretation in terms of a first order phase transition. On the other hand, MF predictions agree with observations in thermal vapour experiments \cite{Carr2013b,Marcuzzi2014c}, but are in conflict with field-theoretical \cite{Maghrebia} and exact numerical results of one-dimensional spin chains \cite{honing13}. In two dimensions, MF and variational approaches predict the emergence of antiferromagnetic phases at strong dissipation \cite{Lee2011a,Hu2013,Weimer2015b}, contradicting corresponding numerical simulations \cite{Hoening2014a}.

\begin{figure}[t!]
\includegraphics[width=\columnwidth]{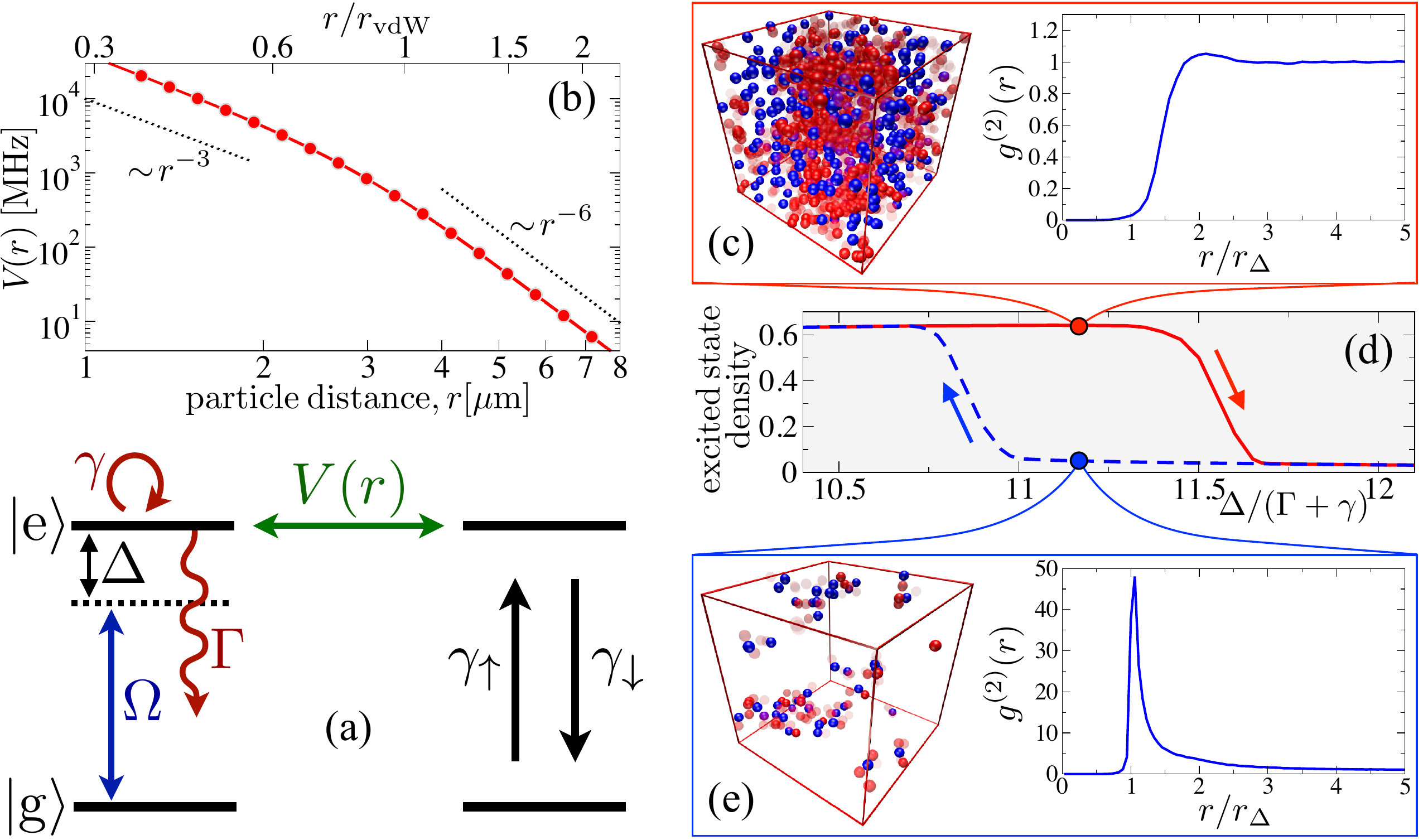}
\caption{\label{fig1} (color online) (a) An ensemble of interacting two-level systems is driven coherently with a coupling strength $\Omega$ and frequency detuning $\Delta$ in the presence of decay and de-phasing with rates $\Gamma$ and $\gamma$, respectively. (b) The potential eq.(\ref{eq:potential}) (solid line) interpolates between different power-laws and accurately describes the actual interaction between excited Rubidium atoms (dots), shown exemplarily for Rb($70S_{1/2}$) atoms. (d) Hysteresis with bistable steady states, whose typical spatial configurations and correlation functions are illustrated in panels (c) and (e). Blue spheres show excited particles, while the opacity of the red dots indicates the excitation rate $\gamma_{\uparrow}$ of particles in state $|{\rm g}\rangle$.
}
\vspace{-0.17in}
\end{figure}

In this Letter we address this problem through numerical studies of the driven-dissipative dynamics in Ising-spin ensembles with power-law interactions. We point out the importance of fluctuations for the topology of the non-equilibrium phase diagram and draw a direct connection to the form of the spin-spin interactions. In particular, bistability cannot occur under the common assumption of pure van der Waals (vdW) interactions, but instead requires a short-distance dipolar potential-core. Such a short-distance behaviour is characteristic for Rydberg atoms [Fig.\ref{fig1}(b)], but typically neglected in theoretical models.
Upon incorporating particle motion we reveal a temperature-driven phase transition to bistable non-equilibrium steady states. In the high-temperature limit we demonstrate a cross-over to MF-behaviour, offering an explanation of thermal-vapour experiments \cite{Carr2013b}.  

We consider a random ensemble of $N$ two-level systems with states $|g_i\rangle$ and $|e_i\rangle$ ($i=1,...,N$) coupled by a laser field with a Rabi frequency $\Omega$ and frequency detuning $\Delta$ [Fig.\ref{fig1}(a)]. The associated unitary dynamics is governed by the Hamiltonian $\hat{H} = \frac{\Omega}{2}\sum_i \left(|e_i\rangle\langle g_i |+|g_i\rangle\langle e_i |\right) -\Delta\sum_i  |e_i\rangle\langle e_i |+\sum_{i< j } V(r_{ij})|e_i e_j\rangle\langle e_i e_j |$, where $V(r_{ij})$ denotes the interaction potential of two particles at positions ${\bf r}_i$ and ${\bf r}_j$ and $r_{ij}=|{\bf r}_j-{\bf r}_i|$. The $N$-body density matrix, $\hat{\rho}$, of the system evolves according to the master equation $\dot{\hat{\rho}} = -i[\hat{H},\hat{\rho}]+\mathcal{L}[\hat{\rho}]$. The Lindblad superoperator $\mathcal{L}[\hat{\rho}]=\sum_{i,\alpha} (L_{i,\alpha} \hat{\rho} L_{i,\alpha}^\dagger -\frac{1}{2}L_{i,\alpha}^\dagger L_{i,\alpha} \hat{\rho}-\frac{1}{2}\hat{\rho}L_{i,\alpha}^\dagger L_{i,\alpha})$ describes one-body decoherence processes. We account for decay of the excited state ($L_{i,0} = \sqrt{\Gamma}|g_i  \rangle \langle e_i|$) with a rate $\Gamma$ and de-phasing of the laser-driven transition ($L_{i,1}=\sqrt{\gamma} |e_i \rangle\langle e_i|$) with a rate $\gamma$.

\begin{figure}[t!]
\includegraphics[width=\columnwidth]{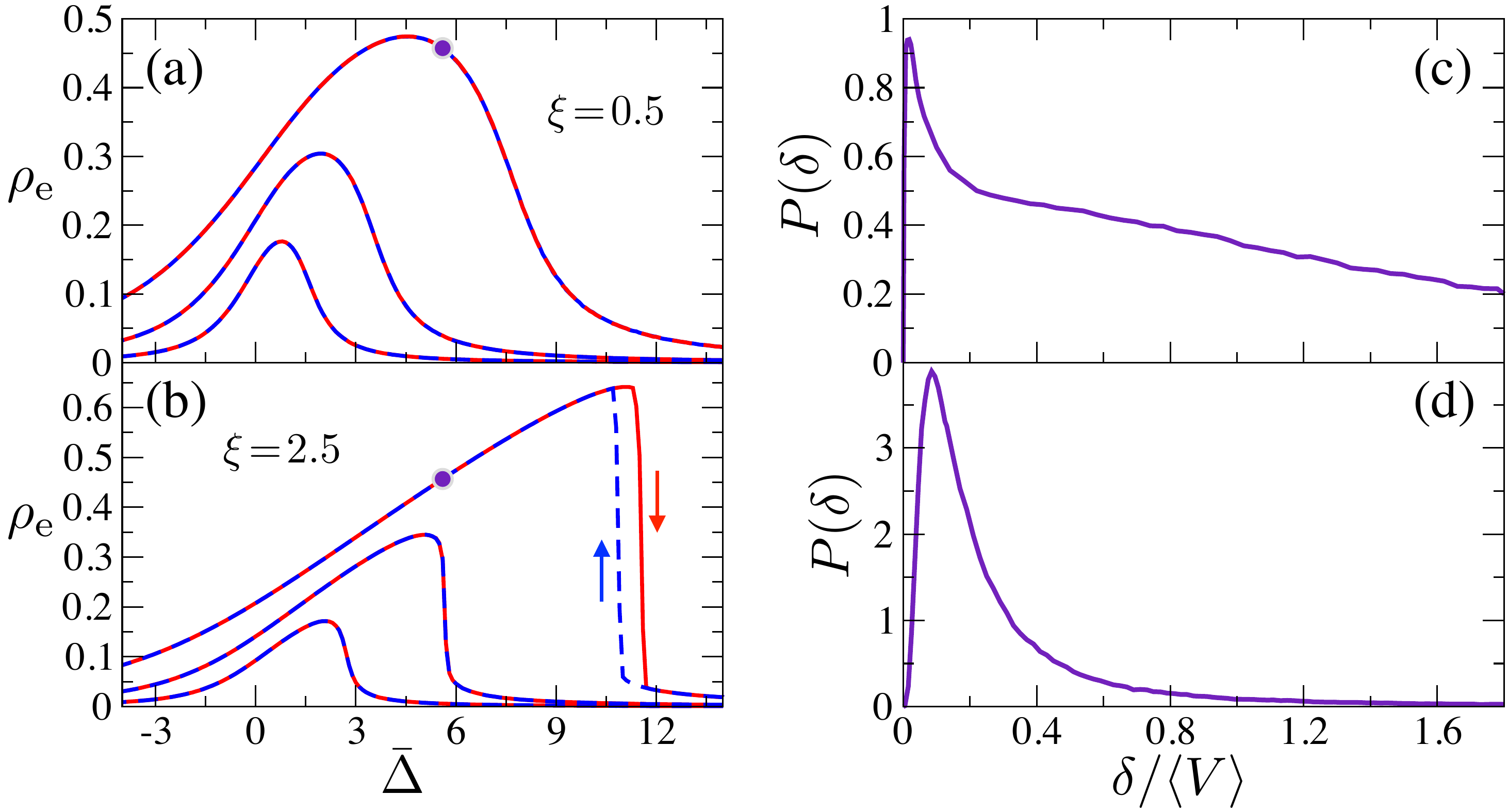}
\caption{\label{fig2}(color online) (a,b) Excitation spectrum for $\rho=10$ and $\Omega=1$, $0.5$ and $0.25$ (from top to bottom), obtained from positive (red solid line) and negative (blue dashed line) frequency scans as indicated by the arrows. (c,d) Distribution of potential fluctuations, $\delta$, for parameters indicated by the dots in panels (a,b).}
\vspace{-0.17in}
\end{figure}

From now on we use dimensionless quantities, scaling time by $\Gamma^{-1}$ and length by the critical radius $r_{\rm b}$, defined as the particle distance at which $V(r_{\rm b})=\Gamma+\gamma$.  We consider a potential \cite{Gallagher2005}
\begin{equation}\label{eq:potential}
 \frac{V(\bar{r})}{\Gamma+\gamma} = \frac{1-\sqrt{1+\xi^6/\bar{r}^6}}{1-\sqrt{1+\xi^6}},
\end{equation}
which features a dipolar potential core ($V\sim1/r^3$) below a distance $r_{\rm vdW}$ and vdW interactions ($V\sim1/r^6$) for $r>r_{\rm vdW}$. Here $\bar{r}=r/r_{\rm b}$ and $\xi=r_{\rm vdW}/r_{\rm b}$ denotes the vdW distance relative to the blockade radius. Since $r_{\rm b}$ defines the typical distance between Rydberg atoms limited by the excitation blockade, the value of $\xi$ characterizes the importances of dipolar interactions. Eq.(\ref{eq:potential}) reproduces the characteristics of Rydberg atom interactions \cite{Saffman2010}, as illustrated in Fig.\ref{fig1}(b) by a comparison to numerical results \cite{bijnen15} for $nS_{1/2}$ states of Rubidium atoms.

Provided that $\Omega/(\Gamma+\gamma)\ll1$ one can adiabatically eliminate the dynamics of the off-diagonal density-matrix elements and obtain a closed evolution equation  for the diagonal $\rho_{{\bf S},{\bf S}}$ \cite{chotia08,Ates2006,Ates2007,Ates2007a}. The matrix elements $\rho_{{\bf S},{\bf S}}$ describe the population of $N$-body configurations $\mathbf{S} \equiv (s_1,\ldots, s_N)$. Here, $s_i$ is an effective spin variable denoting the ground ($s_i=0$) and excited ($s_i=1$) state of the $i^{\rm th}$ particle. Introducing the state vector $\mathbf{S}_i \equiv (s_1,\ldots,1-s_i,\ldots,s_N)$, the resulting master equation  can be written as
\begin{eqnarray}\label{eq:dynamics}
\dot{\rho}_{{\bf S},{\bf S}} &=&-\sum_i \left[ s_i \gamma_\downarrow^{(i)}(\mathbf{S})+(1-s_i)\gamma_\uparrow^{(i)}(\mathbf{S}) \right]\rho_{{\bf S},{\bf S}}\nonumber \\ 
& &+\sum_i \left[ s_i \gamma_\uparrow^{(i)}(\mathbf{S})+(1-s_i)\gamma_\downarrow^{(i)} (\mathbf{S}) \right] \rho_{{\bf S}_i,{\bf S}_i},
\end{eqnarray}
where the single-body (de)excitation rates are given by $\gamma_\uparrow^{(i)} \equiv \bar{\Omega}^2/(1+4\bar{\Delta}_i(\mathbf{S})^2)$ and $\gamma_\downarrow^{(i)} \equiv 1+\gamma_\uparrow^{(i)}$. The rates are determined by two parameters: the scaled Rabi frequency $\bar{\Omega} = \Omega /\sqrt{\Gamma(\Gamma+\gamma)}$ and the scaled frequency detuning $\bar{\Delta}_i({\bf S}) = \Delta_i({\bf S})/(\Gamma+\gamma)$. The latter consists of the laser detuning $\Delta$ and the interaction-induced level shifts from nearby excited particles, $\Delta_i({\bf S}) = \Delta-\sum_{j\neq i}V(r_{ij})s_j$ \cite{Ates2007}. Exact quantum-trajectory simulations for small systems \cite{Sch2014} established the accuracy of this approach for $\Omega\ll\Gamma+\gamma$. Note that this condition does not restrict our parameters, and permits $\bar{\Omega}>1$ if $\gamma>\Gamma$, which is often the case in experiments \cite{Malossi2014}.

The obtained effective master equation can be solved via kinetic Monte Carlo (MC) sampling \cite{MC}. To this end, we randomly sample $N$ particle positions from a cubic volume with periodic boundary conditions and an edge length $L$, chosen to be much larger than $r_{\rm b}$ and $r_{\rm vdW}$. The corresponding dimensionless density $\rho=Nr_{\rm b}^3/L^3$ defines the number of particles within a given blockade volume $r_{\rm b}^3$. To calculate the excitation spectrum, we perform positive and negative scans of the detuning $\bar{\Delta}$ with a corresponding chirp rate $\pm\kappa$. Observables are calculated from an ensemble average over many realizations of particle disorder configurations.

We find two distinct steady states with a low and a high excitation density $\rho_{\rm e}$. The low-density phase corresponds to a dilute gas of excited pairs [Fig.\ref{fig1}(e)], formed by resonant sequential excitation of particles at a distance $r_{\Delta}$ for which the potential $V(r_{\Delta})=\Delta$ compensates the detuning. The correlations in high-density phase [Fig.\ref{fig1}(c)] do not feature strong ordering on the length scale $r_{\Delta}$  and resemble a liquid of repulsive excitations.

\begin{figure}[t!]
\includegraphics[width=\columnwidth]{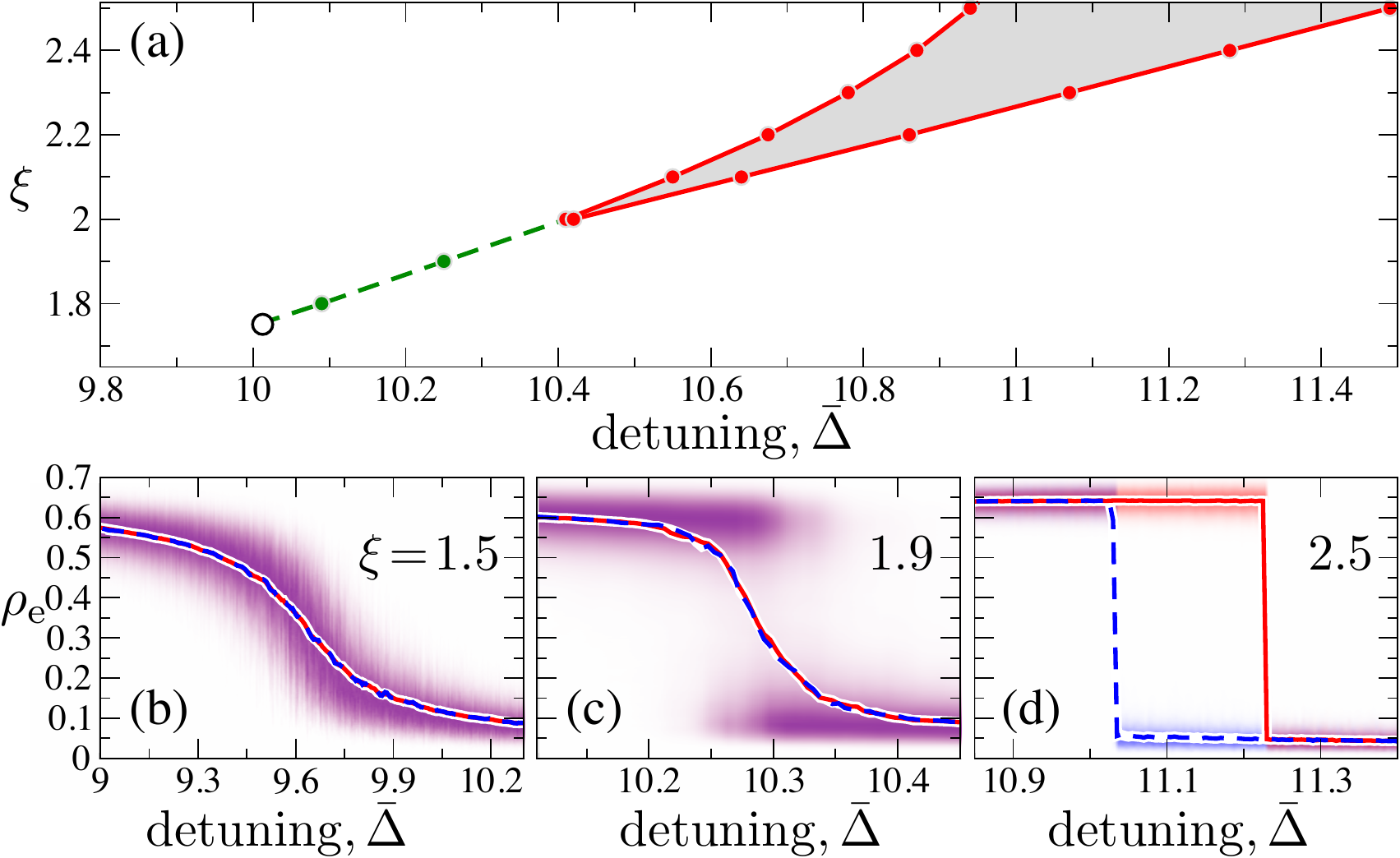}
\caption{\label{fig3} (color online) Phase diagram for $\rho=10$ and $\bar{\Omega}=1$, showing a bistable region (grey area) and first order phase transition (dashed green line) between a low- and high-$\rho_{\rm e}$ phase which ends in a critical point (open circle) at $1.7<\xi<1.8$. (b-d) Excitation density from a single stochastic trajectory. The blue solid and red dashed lines show the average excitation density for different scan directions, while the color shading indicates the corresponding probability distribution of $\rho_{\rm e}$.}
\vspace{-0.17in}
\end{figure}

To investigate the stability of these two phases, we calculate the excitation spectrum for negative and positive scans of the detuning. For a proper choice of parameters, both phases are indeed found to coexist over a finite range of $\bar{\Delta}$ where the excitation density shows hysteretic behaviour, showing qualitative resemblance to MF predictions \cite{Lee2012}. However, in contrast to MF expectations \cite{Marcuzzi2014c}, we find no evidence of bistability for pure vdW interactions ($\xi\rightarrow0$). This is illustrated in Fig.\ref{fig2}(a) and (b) where we show typical excitation spectra for small and large values of the vdW radius. For small $\xi$, the excitation blockade prevents particles from exploring the dipolar region of the interaction potential and one finds a smooth resonance curve with a unique steady state. However, once the short-distance $1/r^3$-behaviour of $V(r)$ starts to become significant the system develops bistable steady states beyond a critical driving strength [Fig.\ref{fig2}(b)].

This behaviour can be understood by considering the effect of the potential form on energy level fluctuations. Spontaneous decay inevitably causes $|e\rangle\rightarrow|g\rangle$ transitions and thereby temporal fluctuations of the corresponding interaction-induced level shifts $\Delta_i$. For $\xi\lesssim1$, the total level shift, $\Delta_i$, of an excited particle typically results from a small number of excitations in close proximity. Hence, a single decay event will cause a substantial change of $\Delta_i$ and disturb the excitation dynamics. The resulting large density fluctuations [Fig.\ref{fig3}(b)] prevent the formation of two distinct phases. For large $\xi$, a large number of excitations within a distance $\lesssim\xi$ collectively contribute to $\Delta_i$, such that potential fluctuations are greatly reduced. To validate this picture, we have traced the microscopic steady state dynamics for two different values of $\xi$ and otherwise identical parameters and average densities, $\rho_{\rm e}$ [dots in Fig.\ref{fig2}(a) and (b)]. By recording the maximum change, $\delta$, of the level shift of excited particles due to a de-excitation, we construct the spectrum of potential fluctuations $P(\delta)$ from the long-time microscopic steady state dynamics. As seen in Fig.\ref{fig2}(c), one indeed finds a broad distribution for $\xi=0.5$ with extended tails well beyond the average potential shift $\langle V\rangle$. On the contrary, for $\xi=2.5$ [Fig.\ref{fig2}(d)] $P(\delta)$ is sharply peaked around small $\delta\ll\langle V\rangle$ and drops rapidly for larger values. It is this strong suppression of fluctuations [Fig.\ref{fig3}(d)] that facilitates the formation of bistable steady states in the limit of large $\xi$. 

\begin{figure}[b!]
\includegraphics[width=\columnwidth]{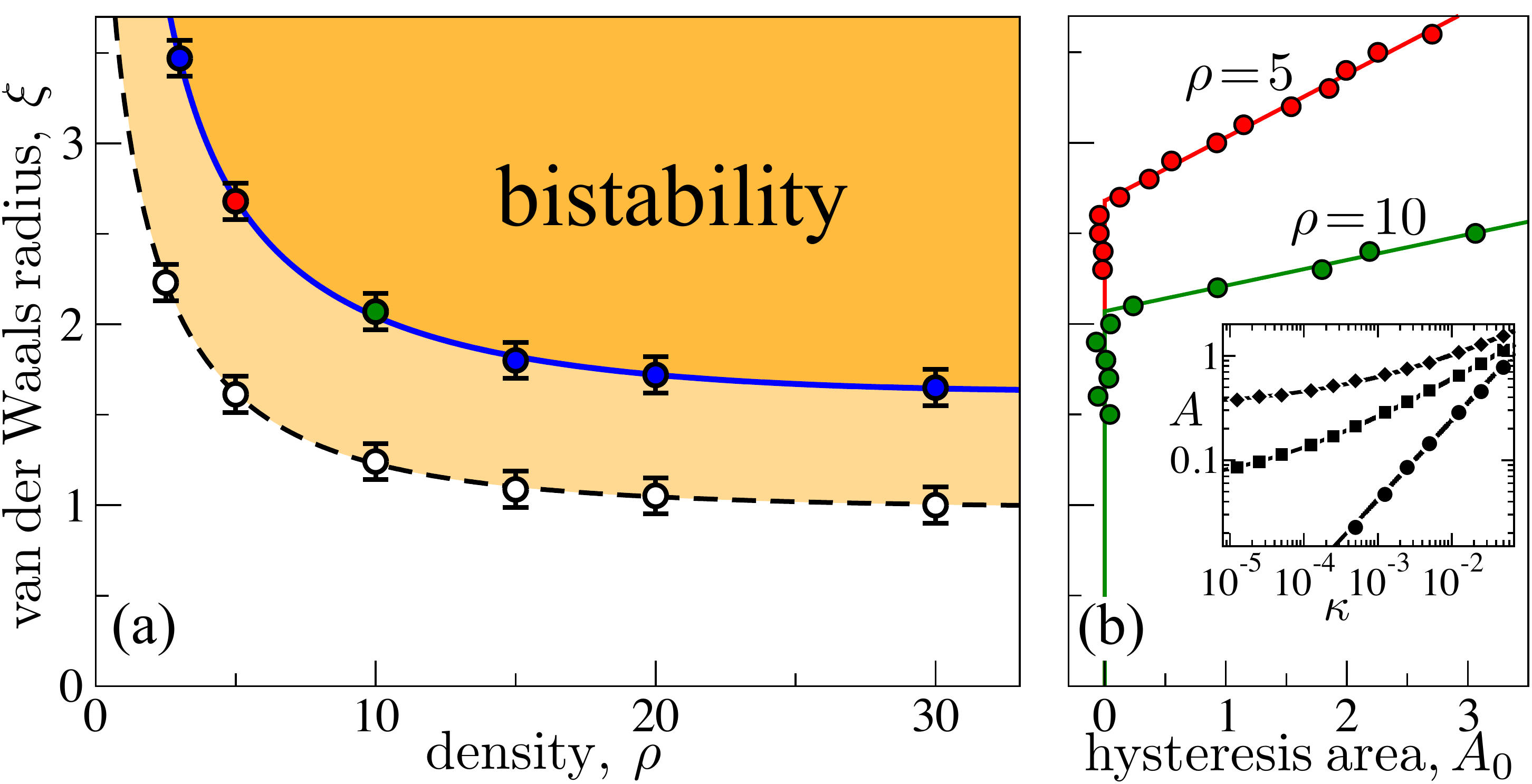}
\caption{\label{fig4} (color online) (a) $\xi-\rho$ phase diagram for $\bar{\Omega}=1$, obtained from MC simulations of frozen particles (solid circles) and eaMF calculations (open circles). (b) shows the static hysteresis area, $A_0$, from the frozen-gas simulations for two different densities. The inset illustrates the typical dependence of $A$ on the chirp rate $\kappa$ for $\rho=10$ and $\xi=1.7$ (circles), $2.1$ (squares) and $2.5$ (diamonds). The lines show a fit to $A=A_0+a\kappa^{-b}$, with free parameters $A_0$, $a$ and $b$.}
\end{figure}

The microscopic steady state dynamics provides further insights about the transition between these two regimes. The non-equilibrium phase diagram shown in Fig.\ref{fig3}(a) reveals a finite region of bistability at large $\xi$ which ultimately closes upon decreasing $\xi$. In between these two limits, the low- and high-density phases, coexisting as long-lived metastable states, are connected by a first order phase transition over a finite range of $\xi$. This transition, generally obscured by the ensemble average over random particle configurations, is revealed by the counting statistics of a single $N$-body trajectory, as demonstrated in Fig.\ref{fig3}(b)-(d) where we show the excitation-density distribution for a single particle configuration at a low chirp rate $\kappa=10^{-8}$. For $1.8\lesssim \xi\lesssim2$ both phases dynamically coexist and yield a \emph{persistent} bimodal counting statistics. The ensemble average of the corresponding transition point yields the first order transition line shown in Fig.\ref{fig3}(a) which ends in a critical point around $1.7<\xi<1.8$. 

For a broader characterization of the conditions leading to bistability we have calculated the asymptotic hysteresis area $A_0$ by extrapolating $A(\kappa)$ to the limit $\kappa\rightarrow0$ [inset of Fig.\ref{fig4}(b)]. Upon changing $\xi$, $A_0$ indicates a continuous transition with a critical exponent $\sim1$ [Fig.\ref{fig4}(b)]. The critical $\xi$ expectedly decreases with particle density [Fig.\ref{fig4}(a)]. Yet, the apparent saturation of the transition line at large densities provides further indication for the absence of bistable behaviour for systems with dominant vdW interactions ($\xi<1)$.

\begin{figure}[t]
\includegraphics[width=\columnwidth]{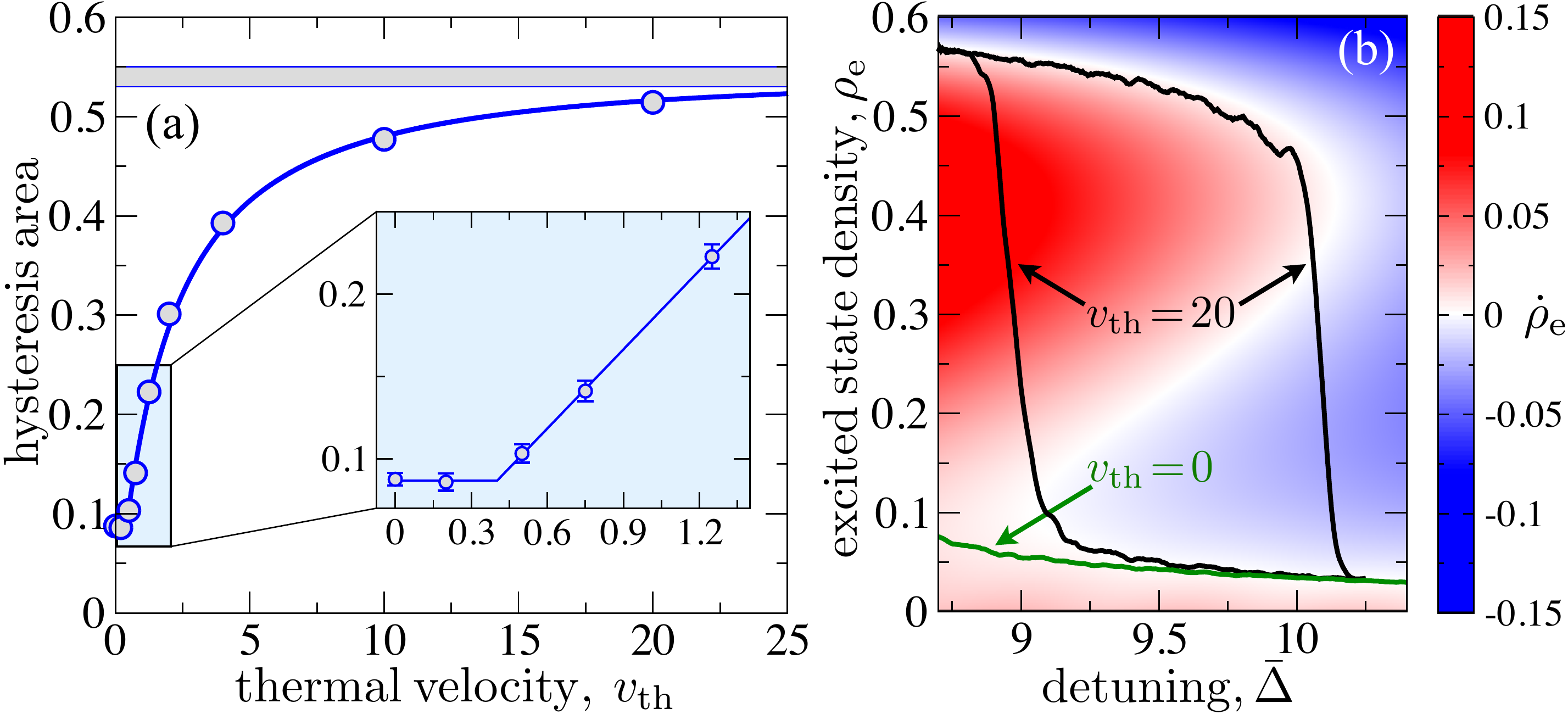}
\caption{\label{fig5} (color online) (a) Hysteresis area as a function of thermal velocity for $\bar{\Omega}=0.8$, $\xi=2$, $\rho=10$, $L=12$ and $\kappa=2.2\cdot10^{-3}$. The data indicates a continuous phase transition around $v_{\rm th}\approx0.4$ and approaches the eaMF prediction shown by the grey horizontal bar. Statistical errors correspond to the symbol size and the height of the bar. The color code in (b) shows $\dot{\rho}_{\rm e}$ from the eaMF approach. The numerical high-temperature simulations ($v_{\rm th}=20$) closely follow the bistable eaMF steady state ($\dot{\rho}_{\rm e}=0$) in contrast to the monostable behavior at $v_{\rm th}=0$.
}
\vspace{-0.17in}
\end{figure}

We have also investigated the average switching time between the two steady states and found the expected power-law divergence upon approaching the transition points in Fig.\ref{fig3}(a). Yet, the corresponding critical exponents strongly depend on $\xi$, which stands in contrast to thermal-vapour experiments \cite{Carr2013b} that observed a universal MF exponent of $0.5$ \cite{MFexp}. To resolve this issue we now consider thermal particle motion which diminishes correlations, and thereby alters the spectrum of fluctuations. For simplicity we neglect inter-particle forces and adopt an ideal gas description with an equilibrium velocity distribution and dimensionless thermal velocity $v_{\rm th}$, measured in units of $r_{\rm b}\Gamma$. Tracking the evolving particle positions, now requires fixed-time-step MC simulations \cite{MC} of the spin dynamics. 

As shown in Fig.\ref{fig5}(a), thermal motion drives a continuous phase transition to bistability.
At high temperatures the hysteresis area saturates to a finite value that can be understood within the following MF treatment. Assuming that rapid thermal motion completely randomizes any spatial excitation structures we can neglect correlations in eq.(\ref{eq:dynamics}) to obtain a closed equation, $\dot\rho_{\rm e}=\bar{\gamma}_\uparrow-(1+2\bar{\gamma}_\uparrow)\rho_{\rm e}$, for the average excitation density. Averaging the microscopic rates $\gamma_{\uparrow}^{(i)}$ over the uncorrelated ensemble yields a closed expression for the MF excitation rate 
\begin{equation}\label{eq:meanfield}
\bar{\gamma}_\uparrow=\frac{\bar{\Omega}^2}{2}\int\! {\rm d}k\:{\rm e}^{-k(1/2+{\rm Re}[f(k)])}\cos\left(k(\Delta+{\rm Im}[f(k)])\right), 
\end{equation}
where $f(k)=k^{-1}\rho_{\rm e}\int 1-{\rm e}^{ikU(r)}{\rm d}{\bf r}$ can be interpreted as the interaction induced line shift [${\rm Im}(f)$] and broadening [${\rm Re}(f)$]. As shown in Figs.\ref{fig5}(a) and (b), our high-temperature results indeed approach this ensemble averaged mean-field (eaMF) limit.
In contrast to corresponding lattice MF models \cite{Marcuzzi2014c,Weimer2015a,Weimer2015b}, the functional $\rho_{\rm e}$-dependence of $\bar{\gamma}_\uparrow$ depends strongly on the shape of the interaction potential. In particular, for $\xi=0$ one finds ${\rm Re}(f)={\rm Im}(f)\propto\rho_{\rm e}\sqrt{k}$, which implies that no phase transition can occur for pure vdW interactions. The numerically obtained eaMF transition line (Fig.\ref{fig4}) demonstrates that this remains true for finite $\xi<1$.

Finally, we put our findings into the context of recent experiments. Ref. \cite{Malossi2014} reports bimodal counting statistics of Rydberg excitations in a cold gas of Rb atoms excited to $70S_{1/2}$ states. The quoted laser linewidth $\gamma/2\pi\approx500~$kHz and Rabi frequencies $\Omega<(\Gamma+\gamma)$ are within the regime of validity of the present theory. The vdW radius of $\xi\approx0.3$ implies that the conditions of \cite{Malossi2014} do not promote bistable steady states. Bimodality at short excitation times, $\tau$, can, however, result from transient relaxation effects \cite{Sch2014}, while for larger $\tau$ dipolar state-mixing induced by black-body radiation \cite{Gallagher1979,Gallagher1980} on a timescale $\tau_{\rm bbr}<\tau$ \cite{Beterov2009,Malossi2014} significantly affects the gas dynamics as observed in other experiments \cite{bbr1,bbr2,bbr3}. Note that the temperature corresponding to a thermal velocity of $r_{\rm b}\Gamma$, for typical values of $r_b\approx11\mu$m and $\Gamma^{-1}\approx200\mu$s \cite{Malossi2014}, can be as low as $\sim30\mu$K such that atom motion can be a factor even in cold-gas experiments.
Consequently, thermal gases \cite{Carr2013b,Urvoy2015} are deep in the high-temperature limit, $v_{\rm th}\gg1$, and their measured excitation dynamics can be understood within the outlined eaMF approach. Importantly, the thermally activated transition to MF behaviour explains the emergence of the dynamical MF exponents \cite{MFexp,Marcuzzi2014c} observed in \cite{Carr2013b}.

In summary, we have investigated driven dissipative spin ensembles with competing power-law interactions. The steady-state of our Master equation (\ref{eq:dynamics}) shares the same MF limit as that \cite{Lee2011a,Marcuzzi2014c} obtained from the exact quantum evolution. Yet it accounts for classical correlations and fluctuations which turn out important for the non-equilibrium physics of such systems. As a striking consequence, the specific shape of the interaction potential was found to play a key role for the non-equilibrium phase diagram despite its general finite-range nature, dropping rapidly as $\sim1/r^6$. The spatial extent of the inner dipolar potential can be tuned by external static \cite{Efield} or microwave \cite{MW} fields, which should permit explorations of the predicted phase diagram (Figs.~\ref{fig3} and \ref{fig4}) in future experiments. We showed that thermal particle motion can drive a transition to bistablity and ultimately causes MF behaviour to emerge. The non-equilibrium phase transition takes place at a surprisingly low temperatures in the $\mu$K- to mK-domain, which should enable its observation in cold atom experiments. The established high-temperature MF limit (eaMF) provides a consistent explanation of recent thermal-vapour experiments \cite{Carr2013b} and, permits future analysis of multi-level atoms and more compex interactions that may occur in such systems \cite{Carr2013b,Urvoy2015}.

\begin{acknowledgments}
We thank Rick van Bijnen, Simon Gardiner, Hendrik Weimer and Igor Lesanovsky for valuable discussions. This work was supported by the EU through the FET-Open Xtrack Project HAIRS, the FET- PROACT Project RySQ, and the Marie-Curie ITN COHERENCE and also by Durham University and the UK EPSRC grants EP/M014398/1 and EP/M013103/1.

\end{acknowledgments}

\end{document}